# Efficient, rapid one-electron photooxidation of chemisorbed polyhydroxyl alcohols and carbohydrates by TiO$_2$ nanoparticles in an aqueous solution.

I. A. Shkrob, [*] M. C. Sauer, Jr., and D. Gozstola

*Chemistry Division, Argonne National Laboratory, Argonne, IL 60439*

**Abstract**

Time-resolved transient absorption spectroscopy has been used to study nanosecond and sub-microsecond electron dynamics in aqueous anatase nanoparticles (*pH*=3-4, 4.6 nm diameter) in the presence of hole scavengers: chemisorbed polyols and carbohydrates. These polyhydroxylated compounds are rapidly oxidized by the holes; 50-60% of these holes are scavenged within the duration of 3.3 ns FWHM, 355 nm excitation laser pulse. The scavenging efficiency rapidly increases with the number of anchoring hydroxyl groups and varies considerably as a function of the carbohydrate structure. A specific binding site for the polyols and carbohydrates is suggested that involves an octahedral Ti atom chelated by the –CH$_2$(OH)-CH$_2$(OH)- ligand. This mode of binding accounts for the depletion of undercoordinated Ti atoms observed in the XANES spectra of polyol coated nanoparticles. We suggest that these binding sites trap a substantial fraction of holes before the latter descend to surface traps and/or recombine with free electrons. The resulting oxygen hole center rapidly loses a CH proton to the environment, yielding a metastable C-centered radical.

___

* To whom correspondence should be addressed: *Tel* 630-252-9516, *FAX* 630-2524993, *e-mail:* shkrob@anl.gov.

# 1. Introduction.

Aqueous anatase (TiO$_2$) nanoparticles find numerous applications in photovoltaics and photocatalysis (e.g., [1,2]). Photoexcitation of the nanoparticles yields electron-hole pairs that rapidly recombine [3-7]; this recombination competes with trapping of the free charges by coordination defects at the nanoparticle surface and lattice defects in the bulk. In this work, we report rapid, efficient scavenging of the holes by polyhydroxyl alcohols and sugars that are chemisorbed at the nanoparticle surface.

It is known from the previous studies that photoexcitation of TiO$_2$ nanoparticles yields at least two kinds of (surface-)trapped holes [8-11], although their exact structure is uncertain. According to EPR [8], both of the hole centers are *O 2p* radicals, such as Ti$^{IV}$-O• [8] and adsorbed hydroxyl radical, OH$_{ads}$ [9]. According to Ishibashi et al. [11], the quantum yield for photogeneration of OH$_{ads}$ is small (ca 0.7% of that for the iodide-oxidizing hole), however, this radical species is very important for photocatalysis, as it is more reactive towards hydrogen donors, such as monohydroxyl alcohols, than other hole centers. For example, Gao et al. demonstrated that only OH$_{ads}$ oxidizes methanol in aqueous TiO$_2$ solutions [10], whereas the predominant oxygen hole center is unreactive towards this substrate and decays via recombination with the electron [12]. Importantly, this conclusion pertains to *aqueous* solutions of the nanoparticles only since time-resolved IR studies of Yamakata at al. [13,14] showed that at the rutile surface exposed to methanol and 2-propanol vapor in vaccum, the holes are scavenged in less than 50 ns, which was the time resolution of their setup. On the other hand, it has been observed that certain polyhydroxylated aliphatic compounds, such as polyvinyl alcohol, are very efficient hole scavengers for TiO$_2$ nanoparticles in an *aqueous* solution (e.g., [15,16]).

In this work, photooxidation of polyols and sugars by aqueous photoexcited TiO$_2$ nanoparticles is studied in a systematic fashion. To this end, nanosecond dynamics of near-IR absorbance from trapped electron were observed following 355 nm laser photoexcitation of the TiO$_2$ nanoparticle solution. Upon the addition of polyhydroxyl alcohols and carbohydrates to the reaction mixture, the survival probability of the electron increases five-fold since scavenging of the hole prevents the decay of the electron via recombination, as the radical product of the scavenging reaction does not recombine with the electron. As shown below, the increase in the survival probability follows the Langmuir-Hinshelwood law. The corresponding scavenging constant uniquely characterizes each hole scavenger, changing with the number of hydroxyl groups and the carbohydrate structure.

**2. Experimental.**

Transient absorbance $\Delta OD_{900}$ from the electron was observed following 355 nm laser photoexcitation of oxygen-saturated $2.4 \times 10^{-4}$ M aqueous solution of 46±5 Å diameter anatase nanoparticles (1400 units of $TiO_2$ per particle) at $pH=4$. This solution was prepared as described in ref. [2] and then aged for a year. The particle size and crystallinity were characterized by TEM [2]. Polyhydroxyl alcohols and sugars of the highest available purity were obtained from Aldrich and used as received.

The 355 nm, 3.3 ns FWHM, laser pulse was derived from the third harmonic of a Nd:YAG laser (Quantel Brilliant). This laser beam was reflected off extra 355 nm dielectric mirrors to remove the traces of the 1st and 2nd harmonic and expanded using a negative lens ($f=-10$ cm) away from the optical cell to uniformly irradiate the cell window masked using a 5.6 mm diameter round aperture. The typical fluence of 355 nm photons through this aperture was 0.05 J/cm$^2$. The analyzing light from a superpulsed Xe arc lamp was passed through a 4 cm water cuvette (not used for > 1 μm detection) followed by a 2-64 glass filter (KOPP, > 650 nm), and then crossed at 45° with the 355 nm laser beam inside a 1.35 mm optical path flow cell with suprasil windows. The OD of the photolysate at 355 nm was 0.8. The transmitted light was focussed, passed through a 3-66 glass filter (KOPP, > 560 nm) and a grating monochromator (SPEX Minimate). For detection at 1-1.35 μm, a set of narrowband (10 nm FWHM) interference filters incremented in 50 nm steps was used. For the detection near 1.05 μm, a 0° dielectric mirror reflecting at 1064 nm was inserted before the detector to beat down the leaking first harmonic of the Nd:YAG laser. A fast Si photodiode (EG&G model FND100Q) biased at -100 V was used to detect the transient absorbance at wavelength < 1.1 μm. For detection at wavelength > 1.1 μm, a Ge photodiode (Germanium Power model GMP566) biased at -10 V or a fast InGaAs photodiode (Germanium Power model GAP520) biased at -5 V were used. The photodiode signal was amplified 10 times using a 1.2 GHz opamp (Comlinear model CLC100) and sampled using a 200 MHz Tektronix TD360 digital oscilloscope or a 1 GHz Tektronix DSA-601 digital signal analyzer. The laser pulse profile was determined using a 1.2 GHz Si photodiode (Hamamtsu model S5792) biased at –12 V. The overall response time of the detection system was ca. 3 ns (for the 350 MHz FND100Q photodiode and 400 MHz bandwidth of the sampler). Absorption signals as small as 100 µOD can be studied with this setup. Light emission from the sample induced by the 355 nm light (which was very minor > 700 nm) was subtracted from the absorbance signal. Typically 8 series of 6 pulses were averaged to obtain the kinetics.

To prevent the buildup of permanent electron absorbance during the laser photolysis of the polyol solutions, the sample was saturated with oxygen and flowed at 1.5 cm$^3$/min during the photolysis (the repetition rate of the laser was 1-2 Hz). A syringe driver was used to flow 10-20 cm$^3$ of the sample during the measurement; after the measurement, the sample was pushed back into the syringe using the excess pressure of the gas, bubbled with oxygen, and reused. No change in the kinetics after 50 such cycles were observed, with or without the hole scavenger. Control experiments using $N_2$-saturated solutions (with no polyols) showed that oxygen does not react with the electron on the microsecond scale (as known from other kinetic measurements [15]), i.e., its only function is to oxidize the electrons between the consecutive laser shots.

Titanium K-edge XANES spectra were obtained using the APS facility at Argonne. The details of the experiment and the analysis are given in refs. [2] and [18].

**3. Results.**

*3.1. Transient absorption spectroscopy and scavenging dynamics.*

The typical kinetics of transient 900 nm absorbance in photoexcited aqueous $TiO_2$ nanoparticles are shown in Fig. 1a,c (trace (i)). Only trapped electron absorbs at this wavelength [3,15,18-20]. Assuming the molar absorptivity of 700±50 M$^{-1}$ cm$^{-1}$ for these electrons at 800 nm [19], ca. 1-1.2 electron-hole pairs per nanoparticle were present in the reaction mixture at the end of 40 mJ/cm$^2$ excitation pulse ($t=0^+$). The quantum yield for the electron, as determined from the prompt absorption signal at $t=0^+$, was ≈0.2, i.e., the recombination within the excitation pulse was relatively inefficient (near-unity quantum yield is commonly assumed for the initial charge separation [5,12]). The observed quantum yield can be compared with the limiting quantum yields for $I_2^-$ formation in aqueous $TiO_2$ solutions containing iodide (an efficient hole scavenger) obtained by other workers: 0.16 [21], 0.42 [16], 0.2 [22], and 0.9 [12]. The latter estimate of 0.9 was obtained at very low fluence of the excitation photons ($< 10^{-9}$ Einstein/cm$^2$); under irradiation conditions similar to ours ($10^{-7}$ Einstein/cm$^2$) the quantum yield was substantially lower (0.3 for $3\times10^{-8}$ Einstein/cm$^2$) [12].

The low-density regime explored in this study is different from most of the previous laser spectroscopy studies (including all of the ultrafast studies of bare $TiO_2$ nanoparticles, e.g. [5,6,7]), in which 10-100 (sometimes, > 1000 [5]) electron-hole pairs were generated per nanoparticle. At such high generation rates, the recombination of trapped electrons and holes occurs on the picosecond time scale, and only a small fraction of the initially generated pairs survives into the nanosecond regime.

Even though the recombination was slowed down due to the low charge density, > 80% of the electrons decayed in less than 100 ns, by intraparticle, single-pair recombination. A small fraction of these electrons decays on a slower time scale, over many decades in time (to at least 1 ms), exhibiting a dispersive, power law kinetics typical of recombination of trapped charges in disordered semiconductors [23] and nanoparticle surfaces [24]. In particular, the "tail" of kinetics shown in Fig. 1c follows $\Delta OD_{900}(t) \propto t^{-\alpha}$ for delay times between 70 ns and 10 µs, with $\alpha \approx 0.46$. This recombination becomes faster on a shorter time scale (Fig. 1c). Between 5 and 300 ns, this kinetic trace can be approximated by a sum of two exponentials with time constants of 6.5 and 73 ns (these time constants slightly vary with the laser fluence). Since the duration of the 355 nm excitation pulse (3.3 ns FWHM) is shorter than either one of the time constants, relatively few (trapped) electrons decay within the duration of the excitation pulse, i.e., the latter integrates over the entire population of these electrons. As there is little decrease in $\Delta OD_{900}$ per decade of time once the power law regime takes over, the ratio of the $t$=300-350 ns absorbance and the prompt absorbance at $t=0^+$ was considered as a fraction of the electrons that "escape" the recombination with the hole on the sub-microsecond time scale.

Qualitatively, the spectral and kinetic changes observed upon the addition of polyols ($C_2$-$C_6$) and carbohydrates to the nanoparticle solution do not depend on the scavenger structure and closely resemble those observed in ref. [25] for glycerol ($C_3$). We address the reader to that work for more detail, as it provides more justification for the analyses described below. It is shown therein that glycerol promptly (< 5 ns) reacts with 40-50% of vis-light-absorbing trapped holes at the nanoparticle surface. Over the first 100 ns, glycerol reacts with most of these light-absorbing holes (not all of these holes are capable of reacting with glycerol on the microsecond time scale); the inferred absorption spectrum of these holes resembles the one observed in photoexcited platinized $TiO_2$ nanoparticles by Bahnemann et al. [17]. The progress of this "slow" hole scavenging can be observed directly from the decay of transient absorbance from the hole at 400-700 nm and the reduction in the bleaching of the 400-900 nm absorbance caused by 532 and 1064 nm laser excitation of trapped electrons (that causes their detrapping and rapid recombination with the holes) [25]. This "slow" hole scavenging (with time constant of 70-100 ns) contributes marginally to the buildup of the electron absorbance in the near-IR because it is much slower than the charge recombination; i.e., most holes which are not scavenged promptly decay by recombination with the electron rather than the "slow" reaction with the glycerol. Importantly, the kinetic changes observed at 900 nm in the presence of glycerol cannot be explained by a change in the light absorption properties of the electron due to the surface modification of the nanoparticle [25]. In particular, the 800-1350 nm spectra obtained from the aqueous and glycerol-containing solutions are identical; also, the overall spectra from the glycerol solutions are the same as those of the electrons generated in aqueous $TiO_2$ solution under the conditions that exclude the formation of holes on the nanoparticles, e.g., by electrochemical reduction of nanocrystalline $TiO_2$ films [20], pulse radiolysis [19], and flash photolysis [4] (of acidic and basic nanoparticle solutions, respectively). Using appropriate weighting coeffcients for the electron and hole spectra, it is possible to reproduce all intermediate transient absorption spectra, for all glycerol concentrations, at any delay time [25].

At 900 nm, the experimental decay kinetics $\Delta OD_{900}^{c}(t)$ of the transient absorbance in the presence of the hole scavenger obey an empirical relation $(t > 0^+)$

$$\Delta OD_{900}^{c}(t) = f(c)\Delta OD_{900}^{c=0}(t = 0^+) + [1 - f(c)]\,\Delta OD_{900}^{c=0}(t) \qquad (1)$$

where $c$ is the molar concentration of the hole scavenger,

$$f(c) \approx f_\infty \, K_h c / (1 + K_h c) \qquad (2)$$

is the fraction of the scavenged holes, and $K_h$ is the scavenging efficiency (Table 1). The same behavior was observed for all wavelengths between 800 and 1350 nm, i.e. across the entire spectral region where only trapped electron absorbs. Note that the time evolution of IR absorbance from electrons at rutile surfaces exposed to methanol and 2-propanol vapor follows the same phenomenology as that given by eq. (1) [13,14].

The empirical formula (1) posits that 900 nm kinetics observed in the presence of a polyol is given by a weighted sum of (i) a flat kinetics and (ii) the kinetics obtained for $c=0$ (trace (i) in Fig. 1a), with the weighting coefficients chosen so that the prompt absorbance $\Delta OD_{900}^c(t = 0^+)$ changes a little (< 5%) in the presence of the scavenger (as seen from Fig. 1a). The flat kinetics originates from the $TiO_2$ nanoparticles in which the hole is scavenged by the polyol (the radical product of the oxidation does not react with the electron), whereas the kinetics whose time profile is given by the scaled trace (i) originates from the $TiO_2$ nanoparticles that still contains an oxygen hole center. To illustrate the applicability of eq. (1), difference traces $\Delta OD_{900}^c(t) - \Delta OD_{900}^c(t = 350 ns)$ for 0 to 1.8 M $D$-arabitol ($C_5$) were normalized at the maximum (Figs. 1b). It is seen from Fig. 1b that these normalized kinetics are very similar. This similarity indicates that all of the holes that can be scavenged by the polyols are scavenged within the duration of the excitation laser pulse. Observe that even at the highest concentration of the hole scavenger, the fraction of the electrons that survive the recombination and persist at longer delay times is 50-60% (Fig. 1a). Rapid as it is, the hole scavenging cannot outcompete electron-hole recombination occurring on the short time scale [3,5,6,7].

Since the prompt 900-1350 nm absorbance increases by less than 2-5% in the presence of the hole scavenger, even at very high concentration of the latter, two-electron oxidation of the polyols is unlikely, at least on the nanosecond time scale. The product of the hole scavenging reaction is a C-centered radical. Some of these radicals, e.g. $CH_2OH$ and $CH(CH_3)_2OH$, are known to "inject" electron into the $TiO_2$ nanoparticles, albeit at a very slow rate [26]. While the "injection" that involves radicals in the water bulk is prohibitively slow (bimolecular rate constants, in terms of nanoparticle concentration, are $2 \times 10^5$ and $3 \times 10^6$ $M^{-1} s^{-1}$ for hydroxymethyl and propan-2-olyl, respectively; see Table 4 in ref. [26]), physisorbed radicals formed in the course of hole scavenging could have reacted faster, and rapid two-electron oxidation is, in principle, possible. However, in such a case, the prompt electron yield would increase in the presence of the scavenger, following the conversion of the holes to electrons. Since such an increase is not observed (Fig. 1a), the prompt two-electron oxidation is not supported by our data. Furthermore, since the decay kinetics of the electron absorbance for $t>0^+$ obey eq. (1) and no growth component due to the delayed conversion of the holes to electrons is apparent, electron injection on the submicrosecond time scale is also ruled out. Once the hole is scavenged by a polyol, the radical product of this reaction neither injects an electron back to the nanoparticle nor recombines with an electron present on the same nanoparticle (on the time scale of our experiment).

This conclusion is in accord with the time-resolved IR study by Yamakata et al. [14] who observed the evolution of electron and C=O stretch absorbances in the laser excitation of the rutile surface exposed to 2-propanol vapor. Whereas the change in the electron decay kinetics (due to hole scavenging) was instantaneous (< 50 ns) [13,14], the C=O signal from acetone at 1700 cm$^{-1}$ appeared very slowly, with a half life of 50 μs (Fig. 3 in ref. 14). Importantly, it was observed that (i) the prompt electron absorbance did not change significantly with the vapor pressure (Fig. 1 in ref. 14), whereas the survival probability was pressure-dependent, and (ii) the hole scavenging was not 100% efficient even at the high vapor pressure. Thus, the polyols adsorbed at the surface of aqueous anatase nanoparticles scavenge holes in much the same way as monohydroxyl alcohols adsorbed at the rutile surface in the gas phase. Furthermore, much the same evolution of the kinetic traces was observed in hole scavenging by iodide [12] and thiocyanate [6,17] (on the time scale from 5 ps to 10 μs).

### 3.2. Scavenging efficiency.

Eq. (2) is illustrated in Figs. 2a,b where the fraction $\Delta OD_{900}^c(t = 315 - 370\,ns)/\Delta OD_{900}^c(t = 0^+)$ is plotted vs. the molar concentration $c$ of the hole scavenger. All these dependencies may be approximated by the Stern-Volmer law, and the scavenging constants $K_h$ thus obtained are given in Table 1. Scheme 1 gives the structure of the sugars.

Whereas all polyols and carbohydrates increased the 900 nm absorbance for t>100 ns, addition of up to 50 vol % of monohydroxyl alcohols (methanol, ethanol, propanol, and 2-propanol) had no effect on the kinetics. This negative result is in agreement with the previous studies of Rabani et al. [12]. With the exception of *D*-arabitol ($C_5$), for linear polyols the constant $K_h$ is proportional to *(n-1)$^2$*, where *n* is the carbon number, i.e., the scavenging efficiency rapidly increases with the number of hydroxyl groups per molecule. Having two hydroxyl groups in the 1,2-position (as in 1,2-ethanediol *vs.* 1,3-propanediol) increases $K_h$. Closing a polyol into a ring considerably decreases $K_h$ (compare *D*-mannitol and inositol). This decrease is accounted for by postulating that in sugars, neighboring hydroxyl groups oriented to the same side of the ring have *higher* scavenging efficiency than same groups oriented to the opposing sides (Scheme 1). The example of *D*-glucal and *D*-galactal (Table 1 and Scheme 1) suggest that the difference in the scavenging efficiency for these two orientations is ca. 2.5 times. As more neighboring hydroxyl groups are added, this difference becomes smaller (compare *D*-glucose and *D*-galactose, Table 1). The highest constants $K_h$ were found for a disaccharide, *D*-lactose, and a furanose, *D*-fructose (Table 1).

Given that hole scavenging is very rapid and structure-specific it is likely that the reaction involves a scavenger molecule that is chemisorbed at the particle surface in a particular way. Recent XANES studies by Rajh and co-workers show that for small aqueous $TiO_2$ nanoparticles, surface titanium atoms change the coordination from octahedral to square pyramidal (with a coordination number of five) [2,18,27]. These undercoordinated sites exhibit enhanced reactivity towards bidentate ligands, such as ascorbic acid, dopamine, and catechol, that bind to these sites and change the coordination of Ti atoms back to octahedral, "repairing" the site [2]. The spectroscopic manifestation of this bidentate binding is a red shift in the absorption spectrum of the coated nanoparticles due to a partial charge transfer from the nanoparticle to the π system of the ligand [2,27]. Since the active site of these ligands, -CH(OH)=CH(OH)-, closely resembles that of the polyols (-$CH_2$(OH)-$CH_2$(OH)-), an experiment was conducted to discern whether the polyols and carbohydrates also bind to undercoordinated titania atoms at the nanoparticle surface.

*3.3. XANES spectroscopy of surface modified nanoparticles.*

The typical Ti K-edge XANES spectra are shown in Fig. 3. To obtain these spectra, polylols were added in a concentration sufficient to obtain the total coverage of Ti atoms at the particle surface, and the solutions were dried. The XANES signal from undercoordinated Ti sites is seen in Fig. 3 as a sharp feature centered at 4.97 keV [2,18]. This feature is present in the $TiO_2$ nanoparticles but absent in crystalline anatase. The polyol- and carbohydrate-coated nanoparticles exhibit a 4.97 keV signal that is intermediate to these two extremes, suggesting that undercoordinated Ti atoms are partially "repaired" by these hole scavengers. The efficiency of this repair appears to decrease with the number of hydroxyl groups in the modifier molecule. At the full surface coverage, ethylene glycol ($C_2$) and glycerol ($C_3$) repair 85% of the undercoordinated surface sites, erythritol ($C_4$) and fructose ($C_6$) repair 54% of these sites, and *D*-arabitol ($C_5$), *D*-glucose ($C_6$) and *D*-glucal ($C_6$) repair 43% of these sites. This observation suggests that extra hydroxyl groups, while increasing the binding to the surface, may not assist in the repair of undercoordinated Ti atoms as efficiently as they do in smaller polyols. In any case, these XANES spectra indicate that polyols and carbohydrates are efficient surface modifiers and that a significant fraction of undercoordinated sites are capped in a manner similar to the related chelating ligands [2,18,27].

**4. Discussion.**

Given the scarcity of structural data on the interaction between polyols and the surface of $TiO_2$ nanoparticles in aqueous solution, we may only speculate on the mechanism for hole scavenging by these hydroxylic compounds.

Very generally, the scavenging constant $K_h$ combines two quantities, not necesserily related: (i) the efficiency of binding to a specific site(s) at the nanoparticle surface and (ii) the efficiency of hole scavenging at this binding site(s). For a 46 Å diameter nanoparticle, the surface area is ca. 6700 Å$^2$ and, assuming that a typical carbohydrate molecule occupies 5-10 Å$^2$, total coverage of the surface of 0.24 mM nanoparticles is achieved at 0.1-0.3 M of the modifier. This estimate, however, assumes that all of the molecules are adsorbed by the nanoparticles, which is unlikely in an aqueous solution. In particular, large critical concentrations $(=K_h^{-1})$ for ethylene glycol and 1,3-propanediol, suggest a partition between free and bound molecules. We believe that rapid increase in the scavenging efficiency with the number of anchoring hydroxylic groups mainly reflects shifting of the equilibrium towards physi- and chemi- sorption of the polyols by the nanoparticle surface, in an aqueous solution.

Once at the surface, a hydroxylic molecule can be chemisorbed. IR spectroscopy provides ample evidence for chemisorption of monohydroxyl alcohols at titania surfaces and the formation of $Ti^{IV}$-O-R groups *in vacuo* (e.g., [13,14]). Polyols can be chemisorbed in a similar manner, and the XANES spectroscopy provides direct evidence that the complexation does occur, at least for undercoordinated titanium sites. The resulting structure may look like that shown in Fig. 4. This optimized-geometry structure was generated using a molecular mechanics modeling program HyperChem 7.5 (obtained from HyperCube, Inc.) that uses MM+ force field. In the octahedral $(-O)_2Ti(OH)_4$ complex shown in Fig. 4, the dihedral O-C-C-O angle in the ligand is close to 60º and the O-Ti-O angle is close to 87º, which results in a distance of 2.73 Å between the bridging oxygens. This distance may be compared to 2.825-2.826 Å and 2.8 Å between the oxygens of neighboring hydroxyl groups in inositol (for *trans-* and *cis-* orientations of these groups, respectively), calculated using the same program. This comparison suggests that polyols and carbohydrates can be readily accommodated into the octahedral complex. Interestingly, the same calculations suggest that *trans-* orientation of the hydroxyl groups in sugars may be favored energetically because the chair conformation of the pyranose is preserved in the complex, whereas binding to *cis-* hydroxyl groups requires a conformational change to a twisted boat.

The above considerations help to visualize the mode of carbohydrate binding to certain sites at the titania surface. We suggest that $K_h$ is mainly determined by the density of such sites in an aqueous solution rather than the efficiency of hole scavenging at these binding sites. Indeed, in the absence of water, rutile surfaces exposed to methanol vapor trap the holes as rapidly and efficiently [13,14] as the polyol-coated aqueous nanoparticles. *In vacuo*, monohydroxyl alcohols (ROH) are chemisorbed at the $TiO_2$ surface as $Ti^{IV}$-O-R groups [13,14]. In acidic aqueous solution, the chemisorption is thermodynamically unfavorable and these alkoxy groups are promptly hydrolyzed. To shift this equilibrium towards a bound structure which is stable in the room-temperature aqueous environment, chelation must occur, and the magnitude of $K_h$ may reflect the efficiency of this specific mode of the complexation.

We turn now to the salient point of the suggested rationale, that is, why chemi- rather than physi- sorbed polyol and carbohydrate molecules serve as efficient hole scavengers (the same conclusion was reached by Yamakata et al. in their study of hole scavenging by monohydroxyl alcohols at the exposed rutile surface [13,14]).

If the molecule is physisorbed, the scavenging reaction involves a *trapped* hole (an oxygen hole center) and a nearby scavenger molecule: the trapping of the hole precedes the scavenging reaction. These trapped holes are, essentially, $Ti^{IV}$-O• radicals [8] and the abstraction of hydrogen by these radicals is probably too slow to compete with the electron-hole recombination that occurs on nano- or subnano- second time scales. By contrast, when the polyols cap the surface sites, free holes generated in the primary photoexcitation process can be trapped at these sites directly, by the bridging oxygens of the complex (Scheme 2). The resulting species rapidly transfers a proton to water and/or adsorbed molecules, yielding a relatively stable, bound $Ti^{(IV)}$-O-•CH- radical. Such surface-linked radicals have been observed in low-temperature EPR studies of photoilluminated $TiO_2$ nanoparticles in frozen aqueous solutions containing alcohols [8]. The fact that these C-centered radicals are bridged to the titanium helps to account for their relative stability towards oxidation. As mentioned above, some C-centered radicals derived from the alcohols, such as $CH_2OH$, decay by injecting the electron into the $TiO_2$ nanoparticle [26]. In the acidic solution, this "injection" is a concerted electron and hydroxyl proton transfer [28]. Since the latter is impossible for a bridged structure, the "injection" is inhibited as it requires the scission of the Ti-O bond.

The verifiable signature of the proposed mechanism is that the hole scavenging occurs on a much shorter time scale than a nanosecond; in fact, it should occur on the picosecond or shorter time scales. Given that 50-60% of the holes can be trapped by the surface Ti complexes, hole scavenging should actively intervene with hole trapping that occurs on these short time scales.

Such an intervention is also needed to account for the constancy of the prompt near-IR absorbance from the trapped electrons in the presence of the hole scavengers.

Since the characteristic life time of the recombination of trapped charges that is observed on the sub-microsecond time scale is 8-10 ns, any fast or ultrafast process that involves rapid conversion of the trapped hole to a radical (incapable of recombination with the electron) would *de facto* increase the prompt yield of the electron at $t=0^+$, by arresting this recombination within the duration of the 355 nm excitation pulse. One way to interpret the observed persistance of the prompt 900 nm absorbance (and the applicability of eq. (1) in general) is to assume that in the low-density regime explored in our study, most of the geminate electron-hole pairs that recombine on sub-nanosecond time scale involve *free* charge carriers. Electron and hole trapping by surface defects (including polyol complexes) slows the charge recombination down, and the nanosecond absorption signal integrates over the slowly recombining pairs of the trapped charges (Scheme 2). The characteristic time for the charge trapping and free carrier recombination is just 100-300 fs [29,30] and 10-20 ps [5,6,7], respectively. Polyol complexes compete for the hole both with other surface defects and the electrons. Perhaps, hole scavenging by the polyols is as rapid as that by adsorbed $SCN^-$ anions at the surface of aqueous suspension of Degussa P-25 powders in the picosecond studies of Colombo and Bowman [6], who used time-resolved diffuse reflectance spectroscopy to observe the evolution of the electron absorbance in the vis (Fig. 3 in ref. [6]). These authors found that most of the hole scavenging occurred within the duration of 310 nm photoexcitation pulse, which was < 100 fs. If the deprotonation of the hole trapped by the polyol complex is slower than its recombination with the short-lived free electron, the overall recombination process is not perturbed and the prompt trapped electron yield does not change.

**5. Conclusion.**

It is shown that carbohydrates and $C_2$-$C_6$ polyols rapidly (< 6 ns) scavenge > 50% of the holes in aqueous anatase nanoparticles. The scavenging efficiency increases with the number of anchoring hydroxyl groups and is sensitive to small structural changes in the carbohydrate. This variation is attributed to the efficiency of chemisorption of the scavenger by the titania surface in the aqueous environment. A specific binding site for polyhydroxylated aliphatic compounds at the nanoparticle surface is suggested that involves an octahedrally coordinated Ti atom that is chelated by neighboring hydroxyl groups of the ligand. This binding accounts for the depletion of pentacoordinated Ti atoms that was observed in the XANES spectra of coated $TiO_2$ nanoparticles. We suggest that these binding sites serve as a trap for short-lived free holes and scavenge a substantial fraction of these holes before the latter descend to other traps and/or recombine with the (free) electrons. The resulting trapped hole rapidly loses a CH proton to the environment, yielding a metastable, $Ti^{IV}$-O- bound C-centered radical.

**Acknowledgement.**

I.A.S. thanks Dr. P. Kamat for stimulating discussions, Dr. Z. Saponjic for the preparation of $TiO_2$ samples, and Dr. L. X. Chen for her help with x-ray spectroscopy. Work performed under the auspices of the Office of Science, Division of Chemical Science, US-DOE under contract number W-31-109-ENG-38. We also gratefully acknowledge financial support from the strategic LDRD grant No. 2000-217-R3I from the Argonne National Laboratory.

**Figure Captions.**

Fig. 1.

Decay kinetics of electron absorbance at 900 nm, following short-pulse 355 nm laser excitation of aqueous anatase nanoparticles. Traces (i) in Figs. 1a and 1c are from the solutions without the hole scavenger. The straight line in Fig. 1c (where the kinetics is given on a double logarithmic scale) corresponds to the power law decay. In the presence of a hole scavenger (*D*-arabitol), the decay kinetics flatten out and the electron absorbance at 300-400 ns increases five-fold, whereas the prompt absorbance increases by 5% only. The molar concentration $c$ of the scavenger was (from bottom to the top, as indicated by an arrow): 0, 0.2, 0.41, 0.084, 0.26, 0.92, and 1.8 mol/dm$^3$. In Fig. 1b, the traces $\delta \Delta OD_{900}(t) = \Delta OD_{900}(t) - \Delta OD_{900}(t = 350\ ns)$ normalized at $t=0^+$ for the same series are plotted on the same graph (different kinetics are shown with different colors). It is seen that all of these normalized kinetics are the same within the experimental error.

Fig. 2.

The ratio $\Delta OD_{900}(t = 315 - 370\ ns)/\Delta OD_{900}(t = 0^+)$ plotted *vs.* molar concentration $c$ of the hole scavenger for (a) polyols and (b) carbohydrates (the same conditions as in Fig. 1). The solid lines drawn through the symbols are the least squares Stern-Volmer fits, eq. (2). The scavenging constants $K_h$ obtained from these plots are given in Table 1. The polyols were (i) ethylene glycol, C$_2$, (ii) glycerol, C$_3$, (iii) *meso*-erythritol, C$_4$, (iv) *D*-arabitol, C$_5$, (v) *D*-mannitol, C$_6$, and (vi) *1,3*-propanediol. The carbohydrates were (i) inositol, (ii) $\alpha$-*D*-glucose, (iii) *D*-galactose, (iv) fructose, (v) *D*-glucal, (vi) *D*-galactal, and (vii) $\alpha$-*D*-lactose. The chemical formulas for these sugars are given in Scheme 1.

Fig. 3

Titanium K-edge XANES spectra of surface modified 46 Å diameter TiO$_2$ nanoparticles (dried aqueous solutions). The feature near 4.97 keV is from pentacoordinated Ti atoms in the chemically-active surface sites. Reference traces (i) and (x) are from uncoated, dried TiO$_2$ nanoparticles and macroscopic anatase crystals, respectively. The surface of the aqueous nanoparticles was modified by addition of (ii) ethylene glycol, (iii) glycerol, (iv) *meso*-erythritol, (v) *D*-arabitol, (vi) *D*-glucal, (vii) α-*D*-glucose, (viii) fructose and (ix) α-*D*-lactose. Addition of these polyhydroxylated compounds results in the reduced 4.97 keV peak from pentacoordinated Ti atoms at the nanoparticle surface due to the "repair" of the site by binding of the –CH$_2$(OH)-CH$_2$(OH)- ligand to these atoms (a possible structure for the resulting complex is shown in Fig. 4).

Fig. 4.

A possible structure for Ti-polyol (in this case, ethylene glycol) complex at the surface of aqueous anatase nanoparticles, as suggested by molecular mechanics calculations (see the text). The Ti atom is octahedrally coordinated by oxygen atoms, two of which are bridging the polyol ligand. The hole is initially trapped by these bridging oxygens; only polyols chemisorbed in the specified fashion are thought to be rapid, efficient hole scavengers.

**Table 1.**

Hole scavenging efficiency $K_h$ for polyhydroxyl alcohols and sugars (eq. (2)) in aqueous solution of $TiO_2$ nanoparticles at 25 °C.

| polyol [a] | $K_h$, $M^{-1}$ |
|---|---|
| *linear chain polyols:* | |
| *1,3*-propandiol | 0.33±0.03 |
| ethylene glycol (C2) | 0.44±0.04 |
| glycerol (C3) | 2.3±0.03 |
| *meso*-erythritol (C4) | 5.85±0.85 |
| *D*-arabitol (C5) | 25±3.6 |
| *D*-mannitol (C6) | 18.5±2.5 |
| *sugars:* | |
| inositol | 9.64±0.7 |
| *α-D*-glucose | 10.1±0.35 |
| *D*-fructose | 33±12 |
| *D*-galactose | 9.8±3 |
| *D*-glucal | 1.5±0.3 |
| *D*-galactal | 3.7±0.9 |
| *α-D*-lactose | 27.1±9.3 |

a) see Scheme 1 for the chemical formulas of these sugars.

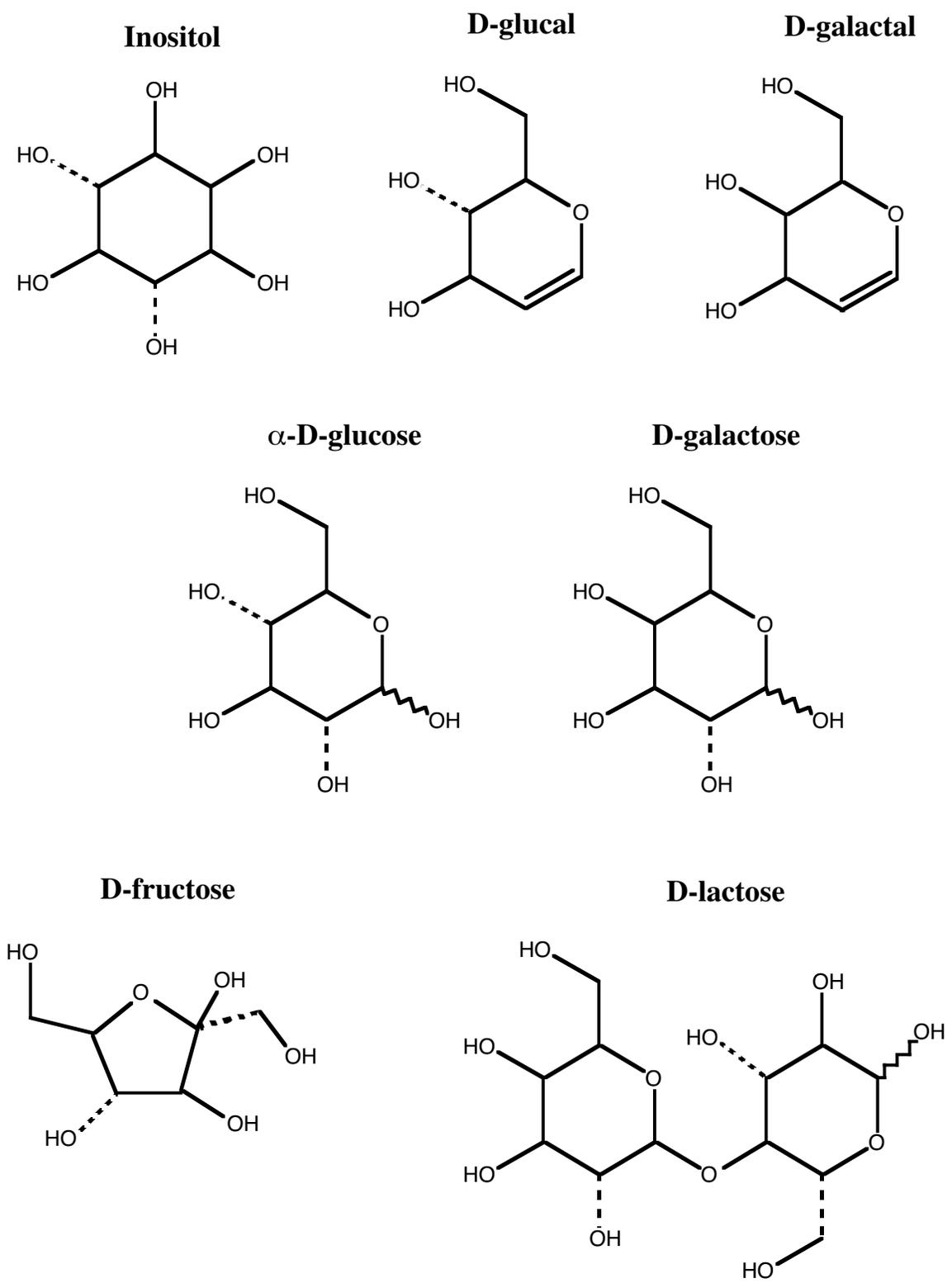

Scheme 1. Chemical formulas for sugars in Table 1.

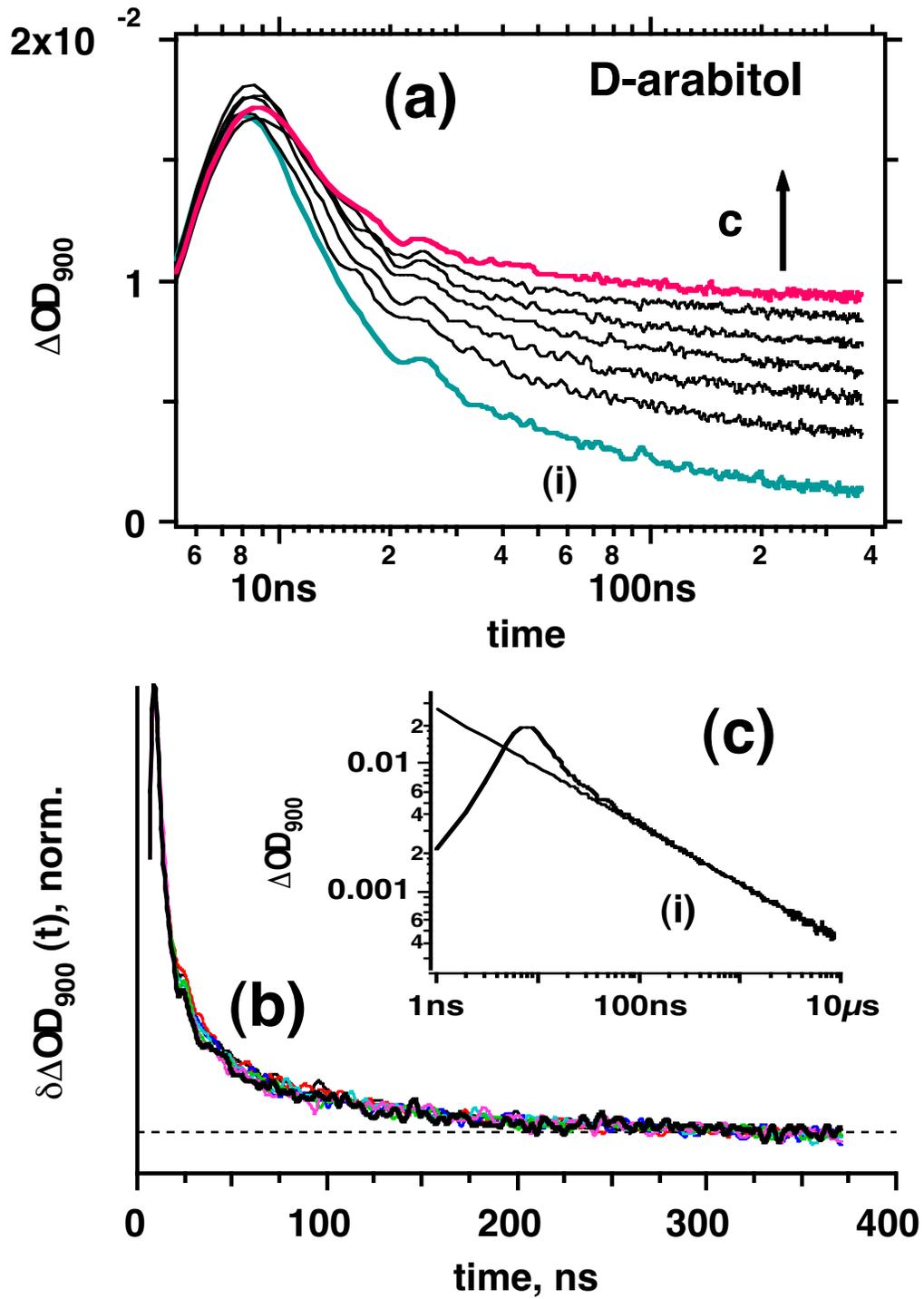

Figure 1.

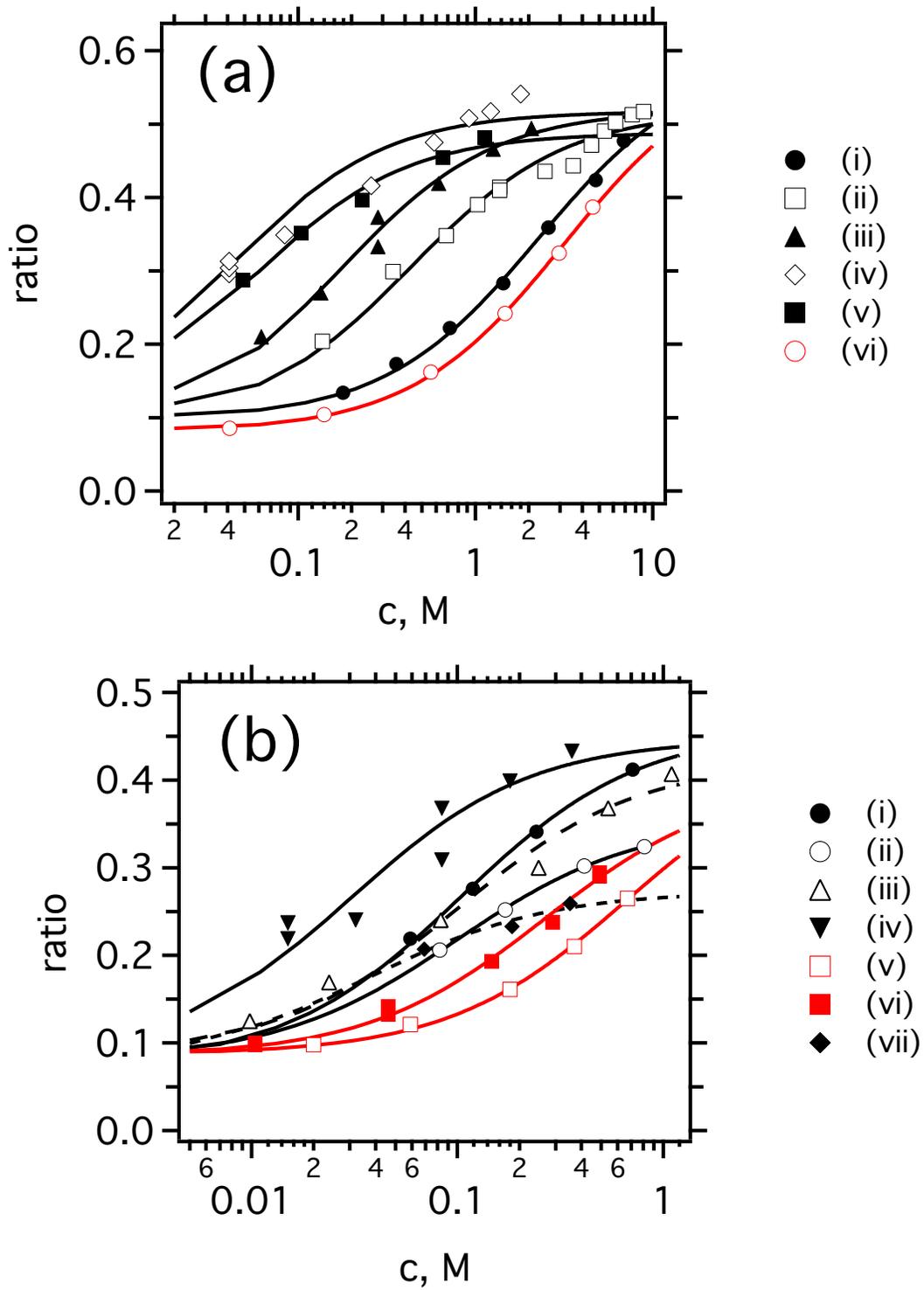

Figure 2.

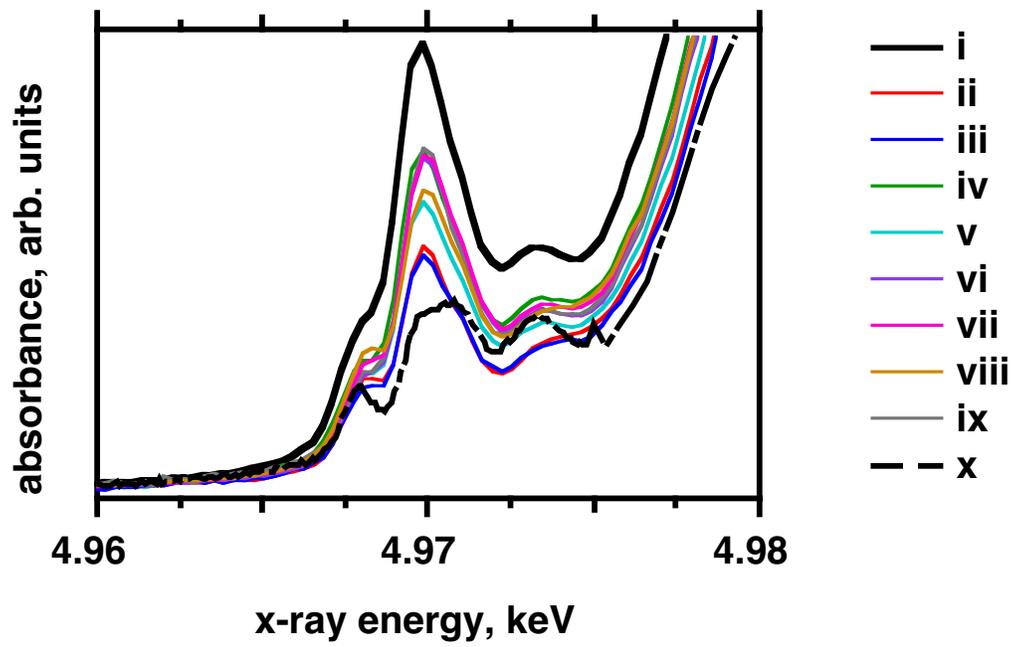

Figure 3.

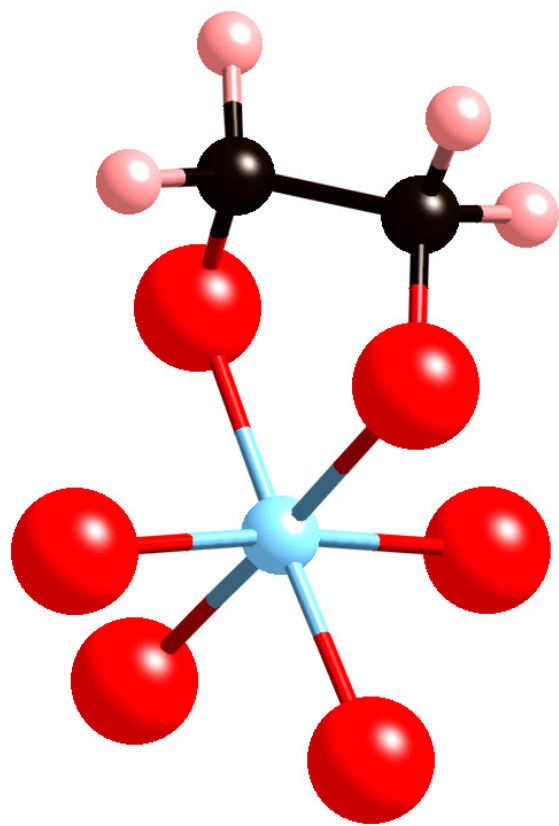

Figure 4.